\documentclass[11pt]{article}
\usepackage{amsmath}
\usepackage{amsfonts}
\usepackage{amssymb}
\usepackage{graphicx}

\def\bea{\begin{eqnarray}}
\def\eea{\end{eqnarray}}
\begin{document}
\begin{center}
\LARGE { \bf Galilean Conformal Algebra in Semi-Infinite Space
  }
\end{center}
\begin{center}
{\bf M. R. Setare\footnote{rezakord@ipm.ir} \\  V. Kamali\footnote{vkamali1362@gmail.com}}\\
 {\ Department of Science, Payame Noor University, Bijar, Iran
}
 \\
 \end{center}
\vskip 3cm

\begin{abstract}
In the present work we considered Galilean conformal algebras
(GCA), which arises as a contraction relativistic conformal
algebras~~($x_i\rightarrow \epsilon x_i$, ~~ $t\rightarrow
t$,~~$\epsilon \rightarrow 0$). We can use the Galilean conformal
symmetry to constrain two-point and three-point functions.
Correlation  functions in space-time without boundary condition
were found in \cite{1}. In real situations there are boundary
conditions in space-time, so we have calculated correlation
functions for Galilean confrormal invariant fields in
semi-infinite space with boundary condition in $r=0$. We have
calculated two-point and three-point functions with boundary
condition in fixed time.
\\

\end{abstract}

\newpage

\section{Introduction}
Recently, there has been some interest in extending the AdS/CFT
correspondence to non-relativistic field theories \cite{a,b}. The
Kaluza-Klein type framework for non-relativistic symmetries, used
in Refs. \cite{a,b,le}, is basically identical to the one
introduced in \cite{d} (see also \cite{e}). The study of a
different non-relativistic limit was initiated in \cite{f}, where
the non-relativistic conformal symmetry was obtained by a
parametric contraction of the relativistic conformal group.
Galilean conformal algebra (GCA) arises as a contraction
relativistic conformal algebras \cite{f,c,7}, where in $d=4$ the
Galilean conformal group is a fifteen parameter group which
contains the ten parameter Galilean subgroup. Beside Galilean
conformal algebra, there is another Galilean algebra, the twelve
parameter schr\"{o}dinger algebra \cite{a,b}. The dilatation
generator in the schr\"{o}dinger group scales space and time
differently, $x_i\rightarrow \lambda x_i$, $t\rightarrow
\lambda^2 t$, but in contrast the corresponding generator in GCA
scales space and time in the same way, $x_i\rightarrow \lambda
x_i$, $t\rightarrow \lambda t$. Infinite dimensional  Galilean
conformal group has been reported  in \cite{c}, the generators of
this group are : ~$L^n=-(n+1)t^n
x_i\partial_i-t^{n+1}\partial_t$,~$M^{n}_i=t^{n+1}\partial_i$
and~$J_{ij}^n=-t^n(x_i\partial_j-x_j\partial_i)$~ for an arbitrary
integer $n$, where  i and  j are specified by the spatial
directions. There is a finite dimensional subgroup of the
infinite dimensional Galilean conformal group which is generated
by ~($J_{ij}^0, L^{\pm 1}, L^0, M_i^{\pm 1}, M_i^0$). These
generators are obtained by contraction ($~t\rightarrow t, ~x_i
\rightarrow \epsilon  x_i, ~\epsilon \rightarrow 0,~ v_i \sim
\epsilon$ ) of the relativistic conformal generators. Several
conformal extensions of the Galilean Lie algebra have been
obtained recently \cite{henk}.
\\ The
representation for finite GCA was found in \cite{m} and \cite{Ne}.
According to the theorem  \cite{n} "Every representation $D(G)$
of a finite group on an inner product space is equivalent to a
unitary representation," the representations of finite  Galilean
conformal group are unitary and we can use this group for
physical applications. For example,  non-relativistic limit of
conformal hydrodynamic describes the small fluctuations from
thermal equilibrium. Whenever  the conformal invariant
dissipative hydrodynamics are considered in nonrelativistic limit,
the relations ~$\partial_{\nu}T^{\mu\nu}=0$~ and
~$T_{\mu}^{\mu}=0$~( The stress-energy tensor of the CFT obeys
these relations ) are converted to incompressible Navier-Stokes
(NS) equation \cite{j}. From reference \cite{h} it can be
inferred that, the Navier-stokes equation for incompressible flow
~($\nabla.\mathbf{v}=0$, $\mathbf{v}$ is velocity of fluid )~ is
covarianted under the infinite GCA. The NS equation for ideal
hydrodynamics (without viscosity) changes to Euler equation. The
finite Galilean conformal group is symmetry of the Euler equation
\cite{c} (see also\cite{k,l}). The gravity dual of finite GCA was
considered in \cite{c,7,i} and the metric with finite 2d GCA
isometry was obtained in \cite{g}. We can use the Galilean
conformal symmetry to constrain correlation function \cite{1}.
Correlation functions of Galilean conformal invariant fields
without any boundary condition were found  in \cite{1}.\\ The
presence of free surfaces or walls in macroscopic systems which
are at the critical point, lead to the large variety of physical
effects. Since, using boundary condition effects is shown to be
very helpful in various branch in physics, the  systems with
boundary conditions have been considered by both theorists
\cite{2} and experimentalists \cite{23456ja}. The situation with
walls or free surfaces opens a new area in condensed matter
physics \cite{1ja}. In reference \cite{4}, the research on
semi-infinite systems which exhibits a non-equilibrium bulk phase
transitions was initiated and  the effects of boundary condition
on direct percolation were considered.\\  Correlation functions
near the boundary are different from other places, and in real
situation there are boundary conditions in space coordinates. By
using some methods in non-relativistic conformal  field theory,
we have obtained an analytic expression for correlation functions
with boundary condition. The situation when a system is in a
predefined initial state and relaxes toward its critical
equilibrium considered as a situation with a boundary condition
at fixed time \cite{4}, for these reasons, in this paper we
calculated two-point and three-point functions in space-time with
boundary condition. The discussion will be exclusively in two
dimensions, but the extension to arbitrary dimension is
immediate. A system with boundary condition at a surface $r=0$
was kept invariant under the transformations were generated by
$[L_{-1},L_{0},L_{1}]$, but space translations, Galilean Boost
transformation and space special Galilean conformal
transformation no longer leave the system invariant. A system
with boundary condition at a fixed time was kept invariant by the
subalgebra $[M_{-1}, M_{0}, M_{1}, L_{0}]$. Correlation functions
with these boundary conditions  are different with the
corresponding results found from Galilean conformal invariance
without boundary condition \cite{1}. If the domain  of space-time
coordinates are infinite in extent (without boundary condition)
and the scalar fields are invariant under the Galilean conformal
(GC) group, the two-point and three-point functions are completely
determined \cite{1}. For example, these results apply to the
calculation of time-delayed correlation functions of
non-relativistic (small viscosity) systems  at equilibrium and at
static critical point. If the space geometry is semi-infinite, a
form of correlation function will be derived, see Eqs. (\ref{12})
and (\ref{27}). These results may be relevant to critical dynamics
close to a surface, see \cite{h14} for an example. when a system
is in a predefined initial state, it can be seen that critical
relaxation towards equilibrium displays  scaling at intermediate
time \cite{4}. (This situation is discribed as a system with
boundary condition in time coordinate.) We  calculate the form of
correlation functions with this condition, see Eq. (\ref{20}).\\
The paper organized as follows. Section 2 is a brief review of
GCA. In section 3,  we calculate the form of correlation functions
 with a boundary condition at surface $r=0$ . Then in
section 4, we calculate the form of two-point function in
space-time with a boundary condition at a fixed time. Then in
section 5, we extend these calculation to the three-point
correlation function. Finally, in section 5, we close by some
concluding remarks.
\\
\section{Representations of  Galilean conformal group}
Galilean conformal  algebras (GCA) was obtained via a direct
contraction of conformal generators. Physically,  this comes from
taking $ t\rightarrow t~~~$, $x_{i}\rightarrow \epsilon
x_{i}~~~$  where $\epsilon\rightarrow 0 $.
 The generators of conformal group are

\begin{eqnarray}\label{1}
  P_{\mu}=-i\partial_{\mu},~~~~~D=-i x^{\mu}\partial_{\mu},~~~~~J_{\mu\nu}=i(x_{\mu}\partial_{\nu}-x_{\nu}\partial_{\mu}),~~~\\
  \nonumber
  K_{\mu}=-i(2x_{\mu}x^{\nu}\partial_{\nu}-x^2\partial_{\mu})~~~~~~~~~~~~~~~~~~~~~~~~~~~~~~~~~~~~
\end{eqnarray}

(where $\mu,\nu=0,1,...,d$).\\
 From the above scaling we obtain the Galilean conformal vector field generators

\begin{eqnarray}\label{2}
 L_{-1}=\partial_{r}~~~~M_{-1}=\partial_{t}~~~L_{0}=(t\partial_{t}+r\partial_{r})~~~J_{ij}=-(x_{i}\partial_{j}-x_{j}\partial_{i})~~~~\\
 \nonumber
 L_{1}=(2tr\partial_{r}+t^2\partial_{t})~~~~~~~~~~~M_1=-t^2\partial_{r}~~~~M_0=-t\partial_{r}~~~~~~~~~~~~~~~~~~~~~~
\end{eqnarray}

$L_{-1},M_{-1},L_0,J_{ij},M_0,L_1,M_1$,~ are  spatial
translations,   time translation, dilatation, rotations, boosts,
time component and space components of special conformal
transformation respectively.  In following, the discussion will be
exclusively in two dimensions, but the extension to arbitrary
dimension is immediate. The above generators obey the following
commutation relations in two dimensions, where define the
Galilean conformal algebras.

\begin{eqnarray}\label{3}
 [L_m,L_n]=(m-n)L_{m+n}~~~\\
 \nonumber
 [L_m,M_n]=(m-n)M_{m+n}~\\
 \nonumber
 [M_n,M_m]=0~~~~~~~~~~~~~~~~~~
\end{eqnarray}
The above symmetries  use to constrain two-point function (see
section (3, 4)). The representations of this group was obtained
in \cite{m}. According to the theorem  \cite{n} "every
representation D(G) of a finite group on an inner product space
is equivalent to a unitary representation", the representations
of $2d$  finite Galilean conformal group  are unitary and it can
be used for physical applications. The representations of
Galilean conformal group are built  by Hausdorff formula \cite{1}.
\begin{eqnarray}\label{4}
 [L_{-1},\phi]=\partial_t\phi~~~~~~~~~~~~~~~~~~~~~~~~~~~~~~~~~~\\
 \nonumber
 [L_0,\phi]=(t\partial_t+r\partial_r+\Delta)\phi~~~~~~~~~~~~~~~~~~\\
 \nonumber
 [L_1,\phi]=(t^2\partial_t+2tr\partial_r+2t\Delta-r\xi)\phi~~~~~
\end{eqnarray}
and
\begin{eqnarray}\label{5}
 [M_{-1},\phi]=\partial_{r}\phi~~~~~~~~~~~~~~~~~~~~~~~~~~~~~~~~\\
 \nonumber
 [M_0,\phi]=(-t\partial_r+\xi)\phi~~~~~~~~~~~~~~~~~~~~~~~\\
 \nonumber
 [M_1,\phi]=(-t^2\partial_r+2t\xi)\phi~~~~~~~~~~~~~~~~~~~
\end{eqnarray}
where $\Delta$ is scaling dimension and $\xi$ is rapidity .

\section{Two-point function in semi-infinite space}
Galilean conformal symmetry is related to massless
non-relativistic systems. Correlations of Galilean-quasiprimary
fields were  studied  without any boundary condition \cite{1}, but
in real systems with boundary conditions, we are interested in
correlation function near the boundary, so we consider the effect
of surface at $r=0$. It is kept invariant under the
transformations were generated by $[L_{-1},L_{0},L_{1}]$, but
space translations, Galilean Boost transformation and space
special Galilean conformal transformation no longer leave the
system invariant. Nevertheless, it is known that Galilean
conformal invariance can be used in analogous situations to
constrain the two-point function \cite{1}. For
Galilean-quasiprimary fields, we require covariance only under
the subalgebra $[L_{-1},L_{0},L_{1}]$.\\ Consider the two-point
function of Galilean-quasiprimary fields

\begin{eqnarray}\label{6}
G=G(r_a,r_b,t_a,t_b)=<\phi_{a}(r_a,t_a)\phi_{b}(r_b,t_b)>
\end{eqnarray}

and we require space points to be in the right half-plane, i.e.
$r_{a},r_{b}\geq0$. Time translation invariance gives
$G=G(r_{a},r_{b}, \tau)$, with $\tau=t_{a}-t_{b}$. From scale
invariance we obtain

\begin{eqnarray}\label{7}
<0\mid[L_0,\phi_a\phi_b]\mid0>=0~~~~~~~~~~~~~~~~~~~~~~~~~~~~~~~~~~~~~~~~~~~~~\\
\nonumber
\Rightarrow \sum_{i=a}^{i=b}(t_{i}\partial_{t_{i}}+r_{i}\partial_{i}+\Delta_{i})G~~~~~~~~~~~~~~~~~~~~~~~~~~~~~~~~~~~~~~~~~~~~~\\
\nonumber =(\tau
\partial_{\tau}+r_{a}\partial_{r_{a}}+r_{b}\partial_{r_{b}}+\Delta)G=0~~~~~~~~~~~~~~~~~~~~~~~~~~~~~
  \end{eqnarray}
where $\Delta=\Delta_{a}+\Delta_{b}$. On the other hand, from the
invariance under the time special Galilean conformal
transformation we have

\begin{eqnarray}\label{8}
<0\mid[L_1,\phi_a\phi_b]\mid0>=0~~~~~~~~~~~~~~~~~~~~~~~~~~~~~~~~~~~~~~~~~~~~~\\
\nonumber \Rightarrow
\sum_{i=a}^{i=b}(t_{i}^{2}\partial_{t_{i}}+2t_{i}r_{i}\partial_{i}+2t_{i}\Delta_{i}-r_{i}\xi_{i})G~~~~~~~~~~~~~~~~~~~~~~~~~~~\\
\nonumber
= ((t_{a}^2-t_{b}^2)\partial_{\tau}+2(t_{a}r_{a}\partial_{r_{a}}+t_{b}r_{b}\partial_{r_{b}})~~~~~~~~~~~~~~~~~~~~~~~~\\
\nonumber
+2(t_{a}\Delta_{a}+t_{b}\Delta_{b}-r_{a}\xi_{a}+r_{b}\xi_{b}))G~~~~~~~~~~~~~~~~~~~~~~~~~~\\
\nonumber
=(\tau^2\partial_{\tau}+2t_{b}(\tau\partial_{\tau}+r_{a}\partial_{r_{a}}+r_{b}\partial_{r_{b}})~~~~~~~~~~~~~~~~~~~~~~~~~~~\\
\nonumber +2(r_a\xi_a-r_b\xi_a)+2\tau r_a\partial_{r_{b}}
+2(t_a\Delta_a+t_b\Delta_b))G~~~~~~~~\\
\nonumber =(\tau^2\partial_{\tau}+2(r_a\xi_a-r_b\xi_a)+2\tau
r_a\partial_{r_{b}}+2\tau\Delta_{a})G=0~~~~~~~
\end{eqnarray}
where in the last equation the scale invariance of $G$ was used.

Now, we make the ansatz

\begin{eqnarray}\label{9}
G(r_a,r_b,\tau)=\tau^{-2\Delta_a}G'(u,v),~~~~~~~u=\frac{r_a}{\tau},~~~~~~~~v=\frac{r_b}{\tau}
\end{eqnarray}
which solves for scale invariance, while Eq. (\ref{8}) gives

\begin{equation}\label{10}
(u\partial_{u}-v\partial_{v}-2(u\xi_a+v\xi_b))G'(u,v)=0,
\end{equation}
Note that, the change of variables in Eq.(\ref{9}) are not
singular, because $\tau=t_a-t_b\neq 0$. The general solution of
Eq.(\ref{10}) is found by the method of characteristics (\cite{2}
and \cite{3}).

\begin{equation}\label{11}
G'(u,v)=\chi (uv) \exp(2(u\xi_a-v\xi_b))
\end{equation}

where $\chi$ is an arbitrary function. The final result is

\begin{eqnarray}\label{12}
G(r_a,r_b,\tau)=\delta_{\Delta_a,\Delta_b}\tau^{-2\Delta_a}\chi_(\frac{r_ar_b}{\tau})\exp(\frac{2}{\tau}(r_a\xi_a-r_b\xi_b))
\end{eqnarray}
  It is clear  that, two-point function  near the
boundary  is different  from other places. We note that
analogously to the conformal result \cite{1}, the scaling
dimension have to agree, while in this case we do not have a
constrain on the rapidity  $\xi_{a}$, since the system is not
space special Galilean conformal invariant. One can use the
relation (\ref{12}) for nonrelativistic  conformal hydrodynamics
with small viscosity near the boundary. Two-point function in the
bulk (out of boundary) for nonrelativistic
conformal  hydrodynamics was found in \cite{1}.  \\

\section{ Two-point function for a non-stationary state }
We now consider a situation with a boundary condition at a fixed
time. Boundary conditions of this type are kept invariant by the
subalgebra $[M_{-1}, M_{0}, $ $M_{1}, L_{0}]$. For example, this
may correspond to the situation when a system is in a predefined
initial state and relaxes toward its critical equilibrium state
\cite{4}. Consider two-point function of quasiprimary fields

\begin{equation}\label{13}
G=G(r_a,r_b,t_a,t_b)=<\phi_{a}(r_a,t_a)\phi_{b}(r_b,t_b)>
\end{equation}

Invariance under space translation implies $G=G(r, t_{a},
t_{b})$  with \\ $r=r_{a}-r_{b}$. We next demand invariance under
Galilean Boost transformation.

\begin{eqnarray}\label{14}
<0\mid[M_0,\phi_a\phi_b]\mid0>=0~~~~~~~~~~~~~~~~~~~~~~~~~~~~~~~~~~\\
\nonumber
\Rightarrow \sum_{i=a}^{i=b}(-t_i\partial_{r_i}+\xi_i)G~~~~~~~~~~~~~~~~~~~~~~~~~~~~~~~~~~~~~~~~~~\\
\nonumber
=(-t_a\partial_{r_a}+\xi_a-t_b\partial_{r_b}+\xi_b)G~~~~~~~~~~~~~~~~~~~~~~~~\\
\nonumber =
(-(t_a-t_b)\partial_r+\xi_a+\xi_b)G=0~~~~~~~~~~~~~~~~~~~~~
\end{eqnarray}

Space special Galilean conformal invariance demands that

\begin{eqnarray}\label{15}
<0\mid[M_1,\phi_a\phi_b]\mid0>=0~~~~~~~~~~~~~~~~~~~~~~~~~~~~~~~~~~\\
\nonumber
\Rightarrow\sum_{i=a}^{i=b}(-t_{i}^2\partial_{r_{i}}+2t_i\xi_i)G=0~~~~~~~~~~~~~~~~~~~~~~~~~~~~
\end{eqnarray}
which gives respectively
\begin{eqnarray}\label{16}
 \xi_a=\xi_b~~~~~~~~~~~~~~~~~~~~~~~~~~~~~~~~~~~~~~~~~~~~~~~~~~
\end{eqnarray}
So, the two-point function reads:
\begin{eqnarray}\label{17}
G=G'(t_a,t_b)\exp(\frac{2\xi_a
r}{t_a-t_b})~~~~~~~~~~~~~~~~~~~~~~~~~~
\end{eqnarray}

In analogy to what was done before, we demand invariance under
scale invariance

\begin{eqnarray}\label{18}
<0\mid[L_0,\phi_a\phi_b]\mid0>=0~~~~~~~~~~~~~~~~~~~~~~~~~~~~~~~~~~~~\\
\nonumber
\Rightarrow \sum_{i=a}^{i=b}(t_i\partial_{t_i}+r_i\partial_i+\Delta_i)G~~~~~~~~~~~~~~~~~~~~~~~~~~~~~~~~~~~~~\\
\nonumber
= (t_a\partial_{t_a}+r_a\partial_{t_a}+\Delta_a+t_b\partial_{t_b}+r_b\partial_{t_b}+\Delta_b)G~~~~~~\\
\nonumber =
(t_a\partial_{t_a}+t_b\partial_{t_b}+r\partial_r+\Delta_a+\Delta_b)G=0~~~~~~~~~~~~
\end{eqnarray}
 we find

\begin{equation}\label{19}
G'(t_a,t_b)=t_{a}^{-(\Delta_a+\Delta_b)}\Phi(\frac{t_a}{t_b})
\end{equation}

where $\Phi$ is an arbitrary function. The final result is

\begin{equation}\label{20}
G(t_a,t_b)=t_{a}^{-(\Delta_a+\Delta_b)}\Phi(\frac{t_a}{t_b})\exp(\frac{\xi
r}{t_a-t_b})
\end{equation}
Note that here we have no condition on the exponents because the
system is not invariant under the time special Galilean conformal
transformation. We are able to use the above relation for
nonrelativistic conformal  hydrodynamics with small viscosity  in
critical equilibrium \cite{4}.\\
Our study in sections 3, 4 was an extension of the work of Henkel
in \cite{2}. This paper has  devoted to the study of the
schr\"{o}dinger symmetry algebra, the maximal symmetry of free
schr\"{o}dinger equations, and one of the things has done there
was considering two-point functions invariant under a sub-algebra
of the schr\"{o}dinger algebra. Here we applied the construction
to the Galilean Conformal Algebra. Our results in Eq. (12) and Eq.
(20) are correspond to the Eq. (3.35) and Eq. (3.44) respectively
in \cite{2}. The difference between these results is the form of
the exponential piece ( $r/t$ in the case of the GCA as opposed to
$r^2/t$ in the schr\"{o}dinger algebra) which is due to the
difference in the form of the dilatation generator.
\section{Three-point function in  space-time with boundary condition}
Three-point function for  Galilean conformal invariant fields in
the bulk without boundary condition was calculated in \cite{1}. In
this section, we calculated three-point function near the
boundary for Galilean conformal invariant fields. Three-point
function in real situation with a boundary condition at fixed
time was calculated. Along lines similar to section (3), we wish
to construct three-point function of three GC-invariant fields
$\phi_{i} (\Delta_i, \xi_i)$ with $i=a, b, c$. As we see in
section (3), the correlation functions near the boundary are kept
invariant under the transformations were  generated by $[L_{-1},
L_0, L_1]$. Consider the three-point function of GC-invariant
fields
\begin{eqnarray}\label{21}
G(r_a,r_b,r_c,t_a,t_b,t_c)=<\phi_a(r_a,t_a)\phi_b(r_b,t_b)\phi_c(r_c,t_c)>
\end{eqnarray}
G is invariant under the time translation which is generated by
$L_0$, so ~$G=G(r_a,r_b,r_c,\tau,\sigma)$ where $\tau=t_a-t_c$ and
$\sigma=t_b-t_c$. From scale invariance one can obtain
\begin{eqnarray}\label{22}
<0\mid[L_0,\phi_a\phi_b\phi_c]\mid0>=0~~~~~~~~~~~~~~~~~~~~~~~~~~~~~~~~~~~~~~~~~~\\
\nonumber
\Rightarrow \sum_{i=a}^{i=c}(t_{i}\partial_{t_{i}}+r_{i}\partial_{i}+\Delta_{i})G~~~~~~~~~~~~~~~~~~~~~~~~~~~~~~~~~~~~~~~~~~~~~\\
\nonumber =(\tau
\partial_{\tau}+\sigma\partial_{\sigma}+r_{a}\partial_{r_{a}}+r_{b}\partial_{r_{b}}+r_{c}\partial_{r_{c}}+\Delta_a+\Delta_b+\Delta_c)G=0
\end{eqnarray}
From the invariance under the temporal non-relativistic special
conformal transformation we have
\begin{eqnarray}\label{23}
<0\mid[L_1,\phi_a\phi_b\phi_c]\mid0>=0~~~~~~~~~~~~~~~~~~~~~~~~~~~~~~~~~~~~~~~~~~~~~~~~~~~~~~~~~~~~~\\
\nonumber \Rightarrow
\sum_{i=a}^{i=c}(t_{i}^{2}\partial_{t_{i}}+2t_{i}r_{i}\partial_{i}+2t_{i}\Delta_{i}-r_{i}\xi_{i})G~~~~~~~~~~~~~~~~~~~~~~~~~~~~~~~~~~~~~~~~~~~~~~\\
\nonumber
= ((t_{a}^2-t_{c}^2)\partial_{\tau}+(t_b^2-t_c^2)\partial_{\sigma}+2(t_{a}r_{a}\partial_{r_{a}}+t_{b}r_{b}\partial_{r_{b}}+t_c r_c\partial_c)~~~~~~~~~~~~~~~~~~~~~\\
\nonumber
+2(t_{a}\Delta_{a}+t_{b}\Delta_{b}+t_c\Delta_c)-r_{a}\xi_{a}-r_{b}\xi_{b}-r_c\xi_c)G~~~~~~~~~~~~~~~~~~~~~~~~~~~~~~~~~~\\
\nonumber
=(\tau^2\partial_{\tau}+\sigma^2\partial_{\sigma}+2t_{c}(\tau\partial_{\tau}+\sigma\partial_{\sigma}+r_{a}\partial_{r_{a}}+r_{b}\partial_{r_{b}}+r_c\partial_c)~~~~~~~~~~~~~~~~~~~~~~~\\
\nonumber -r_a\xi_a-r_b\xi_a-r_c\xi_c+2\tau
r_a\partial_{r_{a}}+2\sigma r_b \partial_{r_b}
+2(t_a\Delta_a+t_b\Delta_b+t_c\Delta_c))G~~~~~~~\\
\nonumber =(\tau^2\partial_{\tau}-r_a\xi_a-r_b\xi_a-r_c\xi_c+2\tau
r_a\partial_{r_{a}}~~~~~~~~~~~~~~~~~~~~~~~~~~~~~~~~~~~~~~~~~~~~~\\
\nonumber +\sigma^2\partial_{\sigma}+2\sigma r_b\partial_{r_b}
+2\tau\Delta_{a}+2\sigma\Delta_b)G=0~~~~~~~~~~~~~~~~~~~~~~~~~~~~~~~~~~~~~~~~~~~
\end{eqnarray}
where in last equation the scale invariance (\ref{22}) of
three-point function $G$ was used. The above equations have this
solution
\begin{eqnarray}\label{24}
G=\delta_{\Delta_a+\Delta_b,\Delta_c}\tau^{-2\Delta_a}\sigma^{-2\Delta_b}G'
\end{eqnarray}
If we introduce $G'=G'_1(r_a,r_c,\tau)G'_2(r_b,\sigma)$ or
$G'=G'_1(r_b,r_c,\sigma)G'_2(r_a,\tau)$, the above equations can
be simplified as
\begin{eqnarray}\label{25}
(\tau^2\partial_{\tau}+2\tau
r_a\partial_{r_a}-r_a\xi_a-r_c\xi_c)G'_1=0\\
\nonumber
(\tau\partial_{\tau}+r_a\partial_{r_a}+r_c\partial_{r_c})G'_1=0~~~~~~~~~~~~\\
\nonumber (\sigma^2+2\sigma r_b\partial_{r_b}-r_b\xi_b)G'_2=0~~~~~~~~~~~~\\
\nonumber
(\sigma\partial_{\sigma}+r_b\partial_{r_b})G'_2=0~~~~~~~~~~~~~~~~~~~~~~
\end{eqnarray}
or
\begin{eqnarray}\label{26}
(\sigma^2\partial_{\sigma}+2\sigma
r_b\partial_{r_b}-r_b\xi_b-r_c\xi_c)G'_1=0~~~\\
\nonumber
(\sigma\partial_{\sigma}+r_b\partial_{r_b}+r_c\partial_{r_c})G'_1=0~~~~~~~~~~~~~~\\
\nonumber (\tau^2+2\tau r_a\partial_{r_a}-r_a\xi_a)G'_2=0~~~~~~~~~~~~~~\\
\nonumber
(\tau\partial_{\tau}+r_a\partial_{r_a})G'_2=0~~~~~~~~~~~~~~~~~~~~~~~~
\end{eqnarray}
The general solution of these equations are found by using the
method of characteristic \cite{n}.
\begin{eqnarray}\label{27}
G=\delta_{\Delta_a+\Delta_b,\Delta_c}(t_a-t_c)^{-2\Delta_a}(t_b-t_c)^{-2\Delta_b}\exp(\frac{r_a\xi_a}{t_a-t_c}+\frac{r_b\xi_b}{t_b-t_c})\\
\nonumber(\chi_{1}(\frac{r_ar_c}{(t_a-t_c)^2})\exp(-\frac{r_c\xi_c}{t_a-t_c})+\chi_2(\frac{r_br_c}{(t_b-t_c)^2})\exp(-\frac{r_c\xi_c}{t_b-t_c}))\\
\nonumber +\Sigma (exchanging~~  ~~ b\leftrightarrow c ~~or~~
a\leftrightarrow c)~~~~~~~~~~~~~~~~~~~~~~~~~~~~~~
\end{eqnarray}
where $\chi_1$ and $\chi_2$ are arbitrary functions. Three-point
function near the boundary is different from other places. In the
following, we calculate three-point function for a situation with
boundary condition at fixed time. Similar to section (4)
three-point function (\ref{21}) is kept invariant by the
subalgebra $[M_{-1}, M_{0}, M_1, L_0]$. Invariance under the
spatial translation which is generated by $M_{-1}$ implies
$G=G(r,s,t_a,t_b,t_c)$ with $r=r_a-r_c$ and $s=r_b-r_c$.
Invariance under  non-relativistic Boost transformation is
demanded
\begin{eqnarray}\label{28}
<0\mid[M_0,\phi_a\phi_b\phi_c]\mid0>=0~~~~~~~~~~~~~~~~~~~~~~~~~~~~~~~~~~\\
\nonumber
\Rightarrow \sum_{i=a}^{i=c}(-t_i\partial_{r_i}+\xi_i)G~~~~~~~~~~~~~~~~~~~~~~~~~~~~~~~~~~~~~~~~~~\\
\nonumber
=(-t_a\partial_{r_a}+\xi_a-t_b\partial_{r_b}+\xi_b-t_c\partial_{r_c}+\xi_c)G~~~~~~~~~~~~~\\
\nonumber =
(-(t_a-t_c)\partial_r-(t_b-t_c)\partial_{s}+\xi_a+\xi_b+\xi_c)G=0~~~
\end{eqnarray}
so, three-point function reads
\begin{eqnarray}\label{29}
G=G'(t_a,t_b,t_c)\exp(\frac{r\xi_c}{t_a-t_c}+\frac{s(\xi_a+\xi_b)}{t_b-t_c})
\end{eqnarray}
Space special Galilean conformal invariance demands that
\begin{eqnarray}\label{30}
<0\mid[M_1,\phi_a\phi_b\phi_c]\mid0>=0~~~~~~~~~~~~~~~~~~~~~~~~~~~~~~~~~~\\
\nonumber
\Rightarrow\sum_{i=a}^{i=c}(-t_{i}^2\partial_{r_{i}}+2t_i\xi_i)G=0~~~~~~~~~~~~~~~~~~~~~~~~~~~~
\end{eqnarray}
which gives
\begin{eqnarray}\label{31}
\xi_c=2\xi_b~~~~~~~~\xi_a=\xi_b
\end{eqnarray}
Finally we demand invariance under  non-relativistic dilatation
\begin{eqnarray}\label{32}
<0\mid[L_0,\phi_a\phi_b\phi_c]\mid0>=0~~~~~~~~~~~~~~~~~~~~~~~~~~~~~~~~~~~~~~~~~~~~~~~~~~~~~~~~~~~~~~\\
\nonumber
\Rightarrow \sum_{i=a}^{i=b}(t_i\partial_{t_i}+r_i\partial_i+\Delta_i)G~~~~~~~~~~~~~~~~~~~~~~~~~~~~~~~~~~~~~~~~~~~~~~~~~~~~~~~~~~~~~~~~~\\
\nonumber
= (t_a\partial_{t_a}+r_a\partial_{r_a}+\Delta_a+t_b\partial_{t_b}+r_b\partial_{r_b}+\Delta_b+t_c\partial_{t_c}+r_c\partial_{r_c}+\Delta_c)G~~~~~~~~~~~~~\\
\nonumber =
(t_a\partial_{t_a}+t_b\partial_{t_b}+t_c\partial_{t_c}+r\partial_r+s\partial_s+\Delta_a+\Delta_b+\Delta_c)G=0~~~~~~~~~~~~~~~~~~~~~~\\
\nonumber \Rightarrow
(t_a\partial_{t_a}+t_b\partial_{t_b}+t_c\partial_{t_c}+\Delta_a+\Delta_b+\Delta_c)G'=0~~~~~~~~~~~~~~~~~~~~~~~~~~~~~~~~~~~~~
\end{eqnarray}
we find
\begin{eqnarray}\label{33}
G'=t_{a}^{-(\Delta_a+\Delta_b+\Delta_c)}\Phi_1(\frac{t_b}{t_c})+t_{b}^{-(\Delta_a+\Delta_b+\Delta_c)}\Phi_2(\frac{t_a}{t_c})+t_{c}^{-(\Delta_a+\Delta_b+\Delta_c)}\Phi_3(\frac{t_a}{t_c})
\end{eqnarray}
where $\Phi_1, \Phi_2, $ and $\Phi_3$ are arbitrary functions. The
general solution  is
\begin{eqnarray}\label{34}
G(r_a,r_b,r_c,t_a,t_b,t_c)=\delta_{\xi_a,\xi_b}\delta_{\xi_c,\xi_a+\xi_b}\exp[\frac{(r_a-r_c)\xi_c}{t_a-t_c}+\frac{(r_b-r_c)(\xi_a+\xi_b)}{t_b-t_c}]~~~~~~~~~~~~~~~~~~~~\\
\nonumber
[t_{a}^{-(\Delta_a+\Delta_b+\Delta_c)}\Phi_1(\frac{t_b}{t_c})+t_{b}^{-(\Delta_a+\Delta_b+\Delta_c)}\Phi_1(\frac{t_a}{t_c})+t_{c}^{-(\Delta_a+\Delta_b+\Delta_c)}\Phi_1(\frac{t_a}{t_c})]~~~~~~~~~~~~~~~~~~~~~~~~~\\
\nonumber +\Sigma (exchanging~~  ~~ b\leftrightarrow c ~~or~~
a\leftrightarrow c)~~~~~~~~~~~~~~~~~~~~~~~~~~~~~~~~~~~~~~~~~~~
\end{eqnarray}
The above result is different from three-point function of the
GCA without boundary condition.

\section{Conclusion}
 We can use finite Galilean conformal group to
constrain two-point and three-point functions. Correlation
functions of Galilean conformal invariant fields in the bulk out
of boundary were found in \cite{1}. Correlation functions near the
boundary are different from other places, and in real situation
there are boundary conditions in space coordinates. When a system
is in an initial state and relaxes toward its critical equilibrium
considered as a situation with a boundary condition at fixed
time.  In this paper we considered two real situations:\\ 1. A
system with boundary in space coordinate ($r=0$) was considered
in sections (3), (5) \\ 2. A system with boundary condition at
fixed time was considered in sections (4), (5).\\ The main results
of this paper are the explicit expressions for  Galilean conformal
invariant correlation functions in a semi-infinite geometry as
given in Eqs. (\ref{12}), (\ref{20}), (\ref{27}) and (\ref{34}).
We calculated two-point function with boundary conditions at
fixed time and surface $r=0$, the form of two-point functions
(\ref{12}), (\ref{20}) obviously are different with the
corresponding results found from Galilean conformal invariance
without boundary condition \cite{1}. The form of three-point
functions are obviously different with corresponding results
found from GC-invariant without boundary condition \cite{1}.

Since Galilean conformal symmetry is related to massless (small
viscosity) non-relativistic systems, one of the applications of
GCA is considering nonrelativistic conformal hydrodynamics  with
small viscosity. One can use the Eqs. (\ref{12}), (\ref{27}) and
(\ref{20}), (\ref{34}) for nonrelativistic conformal
hydrodynamics   with boundary condition in surface $r=0$  and
fixed time respectively.\\
  We know that in 2-dimensional space, there is a so-called exotic central extension of the GCA \cite{du}.
   Martelli and Tachikawa \cite{mt} looked into
 the two-point functions in the bulk, but what happens close to a surface no-one has studied yet, as far as we
 know. We keep this interesting study for our future work in this topic.
\section{ Acknowledgments}
We thank Prof. Malte Henkel for reading the paper and helpful
comments and suggestions.

\end{document}